\newcommand\apj{ApJ}
\newcommand\aj{AJ}
\newcommand\apjl{ApJL}
\newcommand\apjs{ApJS}
\newcommand\aap{A\&A}
\newcommand\mnras{MNRAS}
\newcommand\pasp{PASP}

\newcommand\nat{Nature}

\documentclass[useAMS,usenatbib]{mn2e}

\usepackage{comment,amssymb,graphicx,color}

\title[A VLBI search for AGN in Lyman Break Analogs]{A search for AGN in the most extreme UV-selected
  starbursts using the European VLBI Network} 

\author[R. Alexandroff et al.]{R. Alexandroff$^{1}$\thanks{E-mail: rmalexan@princeton.edu
    (RA)}, R. A. Overzier$^{2}$, Zsolt Paragi$^{3,4}$, Antara Basu-Zych$^{5}$, Tim
  Heckman$^{6}$,
\newauthor
Guinevere Kauffmann$^{7}$, Stephen
  Bourke$^{3}$, Andrei Lobanov$^{8}$, 
Andy Ptak$^{5}$, \newauthor  David Schiminovich$^{9}$\\
$^{1}$Department of Astrophysical Sciences, Princeton University, Peyton Hall-Ivy Lane, Princeton, NJ 08544, USA\\
$^{2}$Department of Astronomy, University of Texas at Austin, C1400, 1 University Station, Austin, TX 78712, USA\\
$^{3}$Joint Institute for VLBI in Europe (JIVE), Postbus 2, 7990 AA Dwingeloo, The Netherlands\\
$^{4}$MTA Research Group for Physical Geodesy and Geodynamics, P.O. Box 91, H-1521 Budapest, Hungary\\
$^{5}$Goddard Space Flight Center, Greenbelt, MD 20771, USA\\
$^{6}$Center for Astrophysical Sciences, Department of Physics and Astronomy, Johns Hopkins University, Baltimore, MD 21218, USA\\
$^{7}$Max-Planck-Institute for Astrophysics, D-85748 Garching, Germany\\
$^{8}$Max-Planck-Institut f\"ur Radioastronomie, Bonn D-53121, Germany\\
$^{9}$Department of Astronomy, Columbia University, New York, NY 10027, USA
}

\begin{document}

\date{}

\pagerange{\pageref{firstpage}--\pageref{lastpage}} \pubyear{2010}

\maketitle

\label{firstpage}

\begin{abstract}
  We have used the European VLBI Network (EVN) to observe a sample of
  Lyman Break Analogs (LBAs), nearby ($z<0.3$) galaxies with
  properties similar to the more distant Lyman Break Galaxies (LBGs).  The study of LBGs 
  may help define the feedback relationship between black holes
  (BHs) and their host galaxies. Previous VLA observations have shown
  that the kpc-scale radio emission from LBAs is dominated by
  starbursts. The main targets of this VLBI experiment were selected
  because they possessed emission-line properties between starbursts
  and Type 2 (obscured) AGN.  Eight targets (three star-forming LBAs,
  four composite LBAs, and one Type 1 AGN) were observed at 5 GHz,
  four of which (one star-forming LBA and three composite LBAs) were also 
  observed at 1.7 GHz.  One star-forming LBA 
  and one composite LBA were detected above 5$\sigma$ at 1.7~GHz
  (only), while the AGN was detected at 5~GHz. In both LBAs, the radio
  luminosity ($L_R$) exceeds that expected from supernovae (remnants)
  based on a comparison with Arp220, Arp229A and Mrk273, by factors of
  $2-8$.  The composite LBA exhibits a compact core emitting around
  10$\%$ of the VLA flux density. The high $T_b$ of $3.5\times10^7$ K and
  excess core $L_R$ with respect to the $L_R/L_X$ relation of radio-quiet
  AGN indicate that this LBA possesses an obscured AGN
  ($M_{BH}\sim10^{5-7}$ $M_{\odot}$). In three other composite LBAs
  detected previously in the X-ray, no radio sources were detected,
  indicating either variability or the presence of an obscured AGN
  below our radio sensitivity.  While weak AGN may co-exist with the
  starbursts as shown in at least one of the LBAs, their contribution
  to the total radio flux is fairly minimal. Our results show that the
  detection of such weak AGN presents a challenge at radio,
  X-ray and optical emission-line wavelengths at $z\sim0.2$,
  indicating the great difficulties that need to be overcome in order
  to study similar processes at high redshift when these types of
  galaxies were common.
\end{abstract}


\section{Introduction}

The nuclei of local galaxies contain supermassive black holes. 
Evidence suggests that these BHs have grown over time following
the mass increase of their host galaxies in accordance with the
$M_{bh}$-$\sigma$ or $M_{bh}$-$M_{bulge}$ relations
\citep{magorrian98,gebhardt00}. Furthermore, feedback between these
BHs and their hosts is an essential element of structure formation
models as it has been proposed that they regulate or suppress star
formation as a result of active galactic nucleus (AGN) activity
\citep[e.g.][]{dimatteo05}. One must thus approach galaxy formation
with both star formation and BH accretion in mind.

Starburst galaxies at high redshift, such as the UV-selected Lyman
Break Galaxies or their dusty counterparts such as the sub-mm
galaxies (SMGs), are the precursors of present day massive galaxies
undergoing a phase of intense star formation
\citep[e.g.][]{adelberger05}. A significant fraction of their stellar
mass is contained in large, dense stellar clumps. It has been
suggested that these giant clumps could host seed BHs that merge in
the center of the galaxy \citep{elmegreen08}, where they can continue
to grow further through other accretion events. It has been shown that
the majority of SMGs at high redshift host an AGN as indicated by
their X-ray emission \citep{alexander05}, while the fraction among
LBGs is a few percent at most
\citep{steidel02,lehmer05,laird06,ouchi08,zheng10}. One possible
explanation for this discrepancy is a delay expected between the
moment of onset of the starburst and the moment that a galaxy is able
to feed its BH, as one has to wait for the strong winds and supernovae
(SN) from the most massive stars to have ceased before gas can fall
onto the BH \citep{norman88,davies07}. This would be consistent with
the relatively young ages of typical LBGs at $z\simeq3-4$ compared to
the dustier and more evolved SMGs.

Although the majority of LBGs do not appear to host luminous AGN, it
is difficult to detect the presence of a low-luminosity AGN (LLAGN),
especially when it is at the center of a powerful starburst.  From
studies in the local universe we know that weak or absent X-ray
emission does not necessarily imply that an accreting BH is
absent. LLAGN have a large range in optical-to-X-ray luminosity
ratios, and many are very weak in hard X-rays due to heavy obscuration
\citep{heckman05a}. Other AGN identification techniques, such as the
commonly used diagnostic tests based on optical emission line ratios
\citep[e.g.][``BPT'' diagrams]{baldwin81} can also yield inconclusive
results. These line ratios are susceptible to, e.g., dust,
metallicity, age and star formation history
\citep{kewley06,yuan10}. New low velocity shock models furthermore
show that the line ratios arising in purely star-forming or
starbursting environments are often indistinguishable from those that
are due to a combination of star formation and some AGN contribution,
both occupying the same ``composite'' region of the BPT diagram
\citep{rich11}. These problems can be partly alleviated by
disentangling the emission from different galaxy components (e.g.,
nucleus vs. outer parts) using integral field spectroscopy, but this
is currently not feasible at high redshifts except for the most
extended galaxies.

We previously selected a rare population of nearby ($z<0.3$) galaxies
that are remarkably similar to LBGs in most of their basic physical
properties. These ``Lyman Break Analogs'' (LBAs) are similar to LBGs
in mass, age, size, metallicity, optical extinction, SFR, and gas
velocity dispersion \citep{heckman05b,hoopes07}. They also have
morphologies similar to those at $z=2-4$
\citep{overzier08,overzier10}, have similar emission line kinematics
\citep{basu09,goncalves10}, and a similar interstellar medium
dominated by strong feedback from massive, clumpy starburst regions
\citep{overzier09,heckman11}. While none of the sources show any
evidence for an unobscured (Type I) AGN (largely by selection), it is
unclear whether some may host obscured (Type II) LLAGN. A small subset
of LBAs falls in the composite region of the BPT diagram. These
sources were also found to have the densest and most massive nuclei
reminiscent of extremely young bulges \citep{overzier09}. An obvious
question to ask is whether we see any evidence for BH growth.

At low intrinsic radio luminosities, the total radio emission of AGN is often dominated by the star formation in their host galaxies \citep[e.g.][]{padovani11, kimball11}.  High-resolution Very Long Baseline Interferometry (VLBI) observations
can be used to search for the elusive LLAGN through their compact,
high brightness temperature radio emission amidst other compact
sources corresponding to radio supernovae (RSN) and supernova remnants
(SNR) frequently seen on pc scales by allowing us to identify excess radio emission above that expected for RSN and SNR models \citep{kewley00,bondi05,gallimore06,perez10,biggs10}.   
Previous VLBI surveys have primarily targeted dusty galaxies (e.g. Luminous or
Ultra-Luminous IR galaxies, (U)LIRGs) in the nearby universe
\citep[e.g.][]{corbett03, parra10}. The UV-selected LBA sample allows us to study
the relationship between nuclear starbursts and BHs in an environment
much more similar to that expected at high redshift.  We present the
results from a pilot project using the European VLBI Network (EVN)
exploring for the first time the compact radio structures in the
nuclei of LBAs. The structure of this Paper is as follows. In \S2 we
present our sample and give details on the 6 and 18 cm radio
observations and data reduction methods. In \S3 we present the main
results. We discuss the implications of our results in \S4, and
conclude with a summary in \S5. Throughout this paper, we assume a
cosmology [$\Omega_M$,$\Omega_\Lambda$,$H_0$] $=$ [0.27,0.73,73.0]
(with $H_0$ in km s$^{-1}$ Mpc$^{-1}$) so that the angular scales at
$z$$=$0.1, 0.2, and 0.3 are $\approx$1.8, 3.2, and 4.3 pc mas$^{-1}$.

\section{Sample, Observations, and Data Reduction}

For this experiment we targeted LBAs that were previously detected at
the $\gtrsim$0.5 mJy level in the Very Large Array (VLA) observations
of \citet{basu07}. Our VLBI sub-sample consists of three sources that
are unlikely to host an AGN, and four sources that may be of
starburst-AGN composite type based on their optical emission line
ratios \citep{overzier09}. For comparison, we also include an eighth source
showing strong evidence for pure AGN emission. We note that this
source has similar far-UV properties as the LBAs but it does not
satisfy the full LBA selection criteria because the latter were
designed to exclude pure AGN. Details on the sources are given in
Table \ref{sources}.

Two separate sets of observations were completed using the European VLBI Network (EVN)
over the course of two years. EVN observations
of J0150+1308, J0213+1259, J0921+4509 and J2103$-$0728 were carried
out at 1.7~GHz on three days between 4--8 June 2008. The participating
telescopes were the Lovell Telescope in Jodrell Bank, the 14-element
Westerbork Synthesis Radio Telescope, Effelsberg, Onsala (25m),
Medicina, Noto, Torun, Urumqi, Shanghai and Cambridge. Baselines
therefore ranged from 198\,km to 8476\,km.  The aggregate bitrate per
telescope was 1024~Mbps using 2-bit sampling and 4s integration
time. There were 8$\times$16~MHz subbands each with 32 channels in
both LCP and RCP polarizations.

\begin{table*}
  \caption{Basic source parameters of our sample. The coordinates are from SDSS. 
    The 1.4~GHz VLA flux densities and radio SFRs are from Basu-Zych et
    al. (2007). Optical line ratios are
    from \citet{overzier09}, $L_{FIR}$ and SFR(UV+IR) from
    \citet{overzier11}, and 2-10 keV X-ray luminosities from
    \citet{jia11}. The radio-derived SFRs agree very well with those
    based on the UV+IR, indicating that the radio emission is
    dominated by star formation.}
\label{sources}
\begin{tabular}{ccccccccccc}
\hline
\hline
$\alpha$ (J2000) & $\delta$ (J2000)   & $z$        &$\frac{[OIII]}{H\beta}$ & $\frac{[NII]}{H\alpha}$  & Type   &  $S^{\rm VLA}_{1.4}$ & SFR$_{1.4}$           & SFR$_{UV+IR}$            & log $L_{FIR}$  & log $L_X$\\
(h m s)                  & ($^{\circ}$ ' ")        &                     &           &                     &                    & (mJy)            & (M$_{\odot}$ yr$^{-1}$)                        & (M$_{\odot}$ yr$^{-1}$) & ($L_\odot$) & (erg s$^{-1}$)    \\
\hline
01:50:28.42 & +13:08:58.1                & 0.147  & 2.29 & 0.21 & SF-LBA          & 1.50 & 30.9     & 57.1  & 11.40     & -- \\
08:15:23.39 & +50:04:14.6                & 0.164  & 0.76 & 0.29 & SF-LBA       & 0.67&17.7              & --      & --          &-- \\
13:53:55.90 & +66:48:00.5                & 0.198  & 1.74 & 0.16 & SF-LBA          & 0.57&22.5           & 26.8   & 11.05     & -- \\
\hline
02:13:48.54 & +12:59:51.4                & 0.219  & 0.78 & 0.81 & COMP-LBA  & 0.90&44.3             & 56.1  & 11.38      & 41.63\\
08:08:44.26 & +39:48:52.3                & 0.091  & 0.58 & 0.66 &  COMP-LBA & $<$0.5& $<$6   & 11.5  & 10.47      & 41.60\\
09:21:59.39 & +45:09:12.4                & 0.235  & 0.74 & 0.50 & COMP-LBA  & 1.41&81.5           & 87.1   & 11.57     & 42.36\\
21:03:58.73 & $-$07:28:02.2             & 0.137  & 0.93 & 0.76 & COMP-LBA  & 3.84&68.1        & 62.4   & 11.36     & 41.68\\
\hline
10:29:59.95 & +48:29:37.9                & 0.232  & 4.17 & 0.45 & AGN & 0.85 &--          & --      &  --          &--\\
\hline
\hline
\end{tabular}
\end{table*}

\begin{figure*}
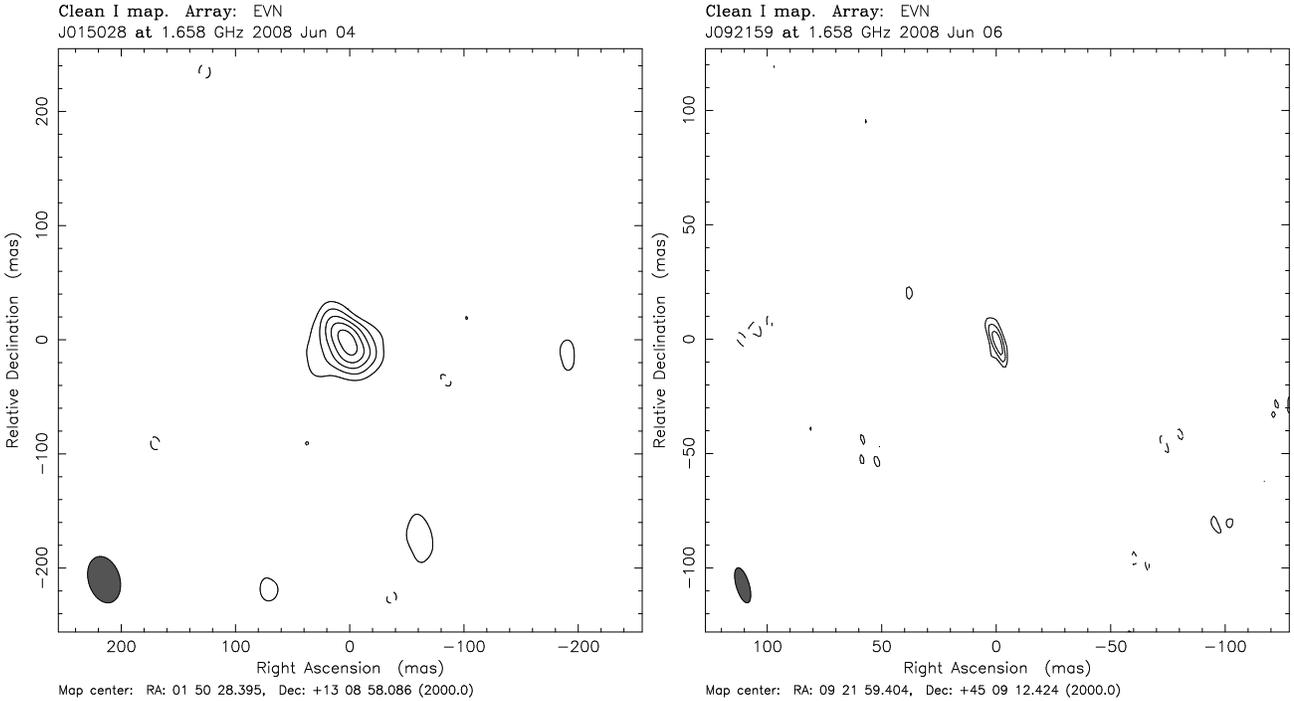

\begin{center}
\includegraphics[angle=0,width=0.48\textwidth,bb= 33 116 582 714,clip]{J015028.ps}
\includegraphics[angle=0,width=0.48\textwidth,bb= 33 116 582 714,clip]{J092159.ps}
\caption{Naturally weighted clean maps of J0150+1308 (left) and
  J0921+4509 (right) at 1.7~GHz. The data for J0150+1308 were
  additionally weighted with a Gaussian taper of 0.5 at a
  $uv$-distance of 10 M$\lambda$. The contour levels are $\pm$1, 1.5,
  2, 2.5 and 3 times the 3-$\sigma$ noise of 45 $\mu$Jy/beam, the
  tapered beam is 41.7$\times$27.5~mas at PA 16.8 degrees, the peak
  brightness is 148 $\mu$Jy/beam. In case of J0921+4509 the contour
  levels are $\pm$1, 1.5, and 2 times the 3-$\sigma$ noise of 45
  $\mu$Jy/beam, the beam is 15.9$\times$5.7~mas at PA 16.3 degrees,
  the peak brightness is 111 $\mu$Jy/beam.}
\label{fig:maps}
\end{center}
\end{figure*}
 
To extend the coherence time, the targets were phase-referenced to
nearby calibrators with separations of 0.5--2 degrees selected from
the VLBA Calibrator
List\footnote{http://www.vlba.nrao.edu/astro/calib/index.shtml}.  The
calibrator positions were accurate to within 0.20 to 1.32~mas. Within
the phase-reference cycle of about eight minutes, 100 seconds was
spent on calibrators and 280 seconds on the targets, and there were
30s gaps included for system temperature measurements. The total
on-source time per target was about 3.5 hours, that would ideally lead
to a thermal noise below 10~$\mu$Jy/beam.

Follow-up EVN observations of an extended sample, now including
J0808+3948, J0815+5004, J1029+4829, J1353+6648 were carried out at
5~GHz on 19-20 March 2010 (see Table~\ref{sources}). A higher
frequency was used in this second run because it was believed to lead
to higher detection rates: the expected flat spectrum emission from an
AGN core would dominate over the steep spectrum emission from
star-forming regions and it was expected that free-free absorption would not 
play a dominant role at this frequency.  The array included the Lovell Telescope,
Yebes (40m), Westerbork, Effelsberg, Onsala (25m), Medicina, Noto and
Torun.  The resulting baselines ranged from 266 to 2279 km. The data
were recorded at 1024~Mbps in a similar fashion to the first
observations (2-bit sampling and a 2-second integration time). There
were 8$\times$16~MHz subbands each with 32 channels in both LCP and
RCP polarizations. Targets were again phase-referenced to calibrators
with angular separations of 0.5-2 degrees.  A phase-reference cycle of
90--300 seconds on the calibrators and targets respectively was
employed. The total on-source time on the targets was between 1--2
hours. The expected rms noise was below $\sim$15~$\mu$Jy/beam allowing
for $>5\sigma$ detections for targets at the level of 100~$\mu$Jy.

The data were correlated by the EVN Data Processor at the Joint
Institute for VLBI in Europe (JIVE).  For data reduction and analysis
we used the NRAO Astronomical Image Processing System (AIPS).  We
followed standard data reduction procedures \citep[e.g.][]{diamond95}: the data
were amplitude calibrated with measured gaincurves and system
temperatures then fringe-fitted and bandpass calibrated using the compact
calibrator sources.  Anomalous solutions were manually flagged in AIPS
using the task $snedt$ to view calculated rate and delay
solutions. Subband 8 for the Lovell telescope and subband 7 for Torun
were completely flagged in the second set of observations using task
$uvflg$ due to poor bandpass solutions. The calibrated target data
were averaged in time and frequency and then exported to UVFITS
format. Additional manual flagging was carried out in Difmap
\citep[][]{shepherd94}.  For the second set of data collected on March
2010, amplitude calibration was verified on several compact
calibration sources (J0209+1352, J0808+4950, J1344+6606) in Difmap.
The sources were cleaned using repeated calls to $clean$ and $selfcal$
and then the amplitude corrections were examined using $gscale$. It
was noted that across all calibrator sources all Torun amplitudes were
off by a factor of 1.7 while Noto subband 7 amplitudes were off by a
factor of 1.5 and subband 8 amplitudes were off by a factor of 1.7.
Systematic corrections were made to these subbands using the AIPS task
$clcor$ and then the data was re-exported to UVFITS format. Possible
sources of such systematic errors include faulty $T_{\rm sys}$
measurements due to RFI usually present across most subbands.

Dirty maps were formed with natural weighting (uvw 0, -2) using the
$mapplot$ command in Difmap. For each target source the peak
brightness and rms noise was determined.  The results are listed in
Table~\ref{results}.  For the two safe detections we show clean maps
in Figure \ref{fig:maps}.

\begin{table*}
  \caption{Measured value of rms noise, peak brightness and signal to
    noise ratio of the peaks for our eight targets (four targets were
    observed both at 1.658 and 4.990~GHz). Upper limits are given for non-detections. Note that because both observations used a different array configuration and the L-band run included Chinese and South African telescopes the resolution from the two runs was quite similar.
  }
\label{results}
\begin{tabular}{ccccccc}
\hline
\hline
Source       & $\nu$ & noise     & peak      & S/N & beam & $T_{\rm b}$\\
             &  (GHz)  & ($\mu$Jy/b) & ($\mu$Jy/b) &     & (mas) & ($10^5$K)\\
\hline
J0150+1308   & 1.658 & 15 &    149 &    9.9 & 8.6$\times$6.2 & $>$32.0\\
J0213+1259   & 1.658 & 14 &  $<$54 & $<$3.9 & 5.9$\times$7.8 & $<$5.0\\
J0921+4509   & 1.658 & 15 &    113 &    7.5 & 5.7$\times$15.9 & $>$1.1\\
J2103$-$0728 & 1.658 & 17 & $<$73 & $<$4.3 & 6.3$\times$10.2 & $<$5.4\\
\hline
J0150+1308   & 4.990 & 26 & $<$111 & $<$4.3 & 6.1$\times$8.5 & $<$1.1\\
J0213+1259   & 4.990 & 26 & $<$115 & $<$4.4 & 6.1$\times$8.7 & $<$1.2\\
J0808+3948   & 4.990 & 35 & $<$162 & $<$4.6 & 5.2$\times$10.1 & $<$1.7\\
J0815+5004   & 4.990 & 31 & $<$117 & $<$3.8 & 5.1$\times$9.3 & $<$1.1\\
J0921+4509   & 4.990 & 25 & $<$100 & $<$4.0 & 5.4$\times$8.1 & $<$1.1\\
J1029+4829   & 4.990 & 24 &    127 &    5.3 & 7.1$\times$9.4 & $>$1.2\\
J1353+6648   & 4.990 & 25 &  $<$98 & $<$3.9 & 5.5$\times$8.3 & $<$1.0\\
J2103$-$0728 & 4.990 & 27 & $<$111 & $<$4.1 & 7.1$\times$10.2 & $<$0.8\\
\hline
\hline
\end{tabular}
\end{table*}

\section{Results}

For the 1.7~GHz data two sources, J0150+1308 (star-forming LBA) and
J0921+4509 (composite LBA) were detected above 5$\sigma$.  J0150+1308
is located at ($\alpha$,$\delta$)$_{J2000}$ of
(01:50:28.39531,+13:08:58.0833). J0921+4509 is located at
(09:21:59.40425,+45:09:12.4232).  Both have a positional error of
approximately 1~mas.  For the 5.0~GHz data only one source was
detected above 5$\sigma$, J1029+4829 (the AGN) while the two sources
detected at 1.7~GHz were not. J1029+4829 is located at
(10:29:59.95930,+48:29:38.0012) with a positional error of
approximately 2~mas.  Real detections should exhibit the synthesized
beam pattern in their images due to the beam pattern's large effect on the noise in
any image, although this cannot be seen at very low signal-to-noise
ratios. The two LBAs detected at 1.7~GHz exhibit this beam pattern in
their dirty images suggesting they are safe detections. The strong
detection of these two sources at 1.7~GHz demonstrates the success of
our phase-referencing efforts.  We believe that the sources that
remained undetected at 1.7~GHz are diffuse radio sources which lack a compact
core detectable on the scale of our observations.  Applying a $uv$-taper to downweight 
longer baselines, testing the effects of possibly over-resolving the flux, did 
not lead to a detection thus strengthening our reasoning that these are indeed faint sources.

At 5~GHz one likely detection just above 5~$\sigma$ was made. For the
source J1029+4829 (AGN) the peak brightness location did not vary when
applying a strong UV-taper or when removing baselines or observation
times suggesting a probable detection. Observations with a greater
sensitivity would be needed to confirm this detection. The two sources
detected at 1.7~GHz were not detected at 5~GHz. This result could be
explained in several possible ways.  One option is that our
phase-referencing did not work well. Alternatively, the sources did
not have compact emission well exceeding the 100$\mu$Jy level so the
sources were completely resolved in this observation. Lastly, it is
possible that they had spectra of ($\alpha \leq -0.1,
S\propto\nu^{\alpha}$) calculated using the 5~$\sigma$ upper limit.

For the detected sources, parameters were obtained by fitting point
and/or circular-Gaussian model components to the $uv$-data using the
$modelfit$ command in Difmap. The results are summarized in
Table~\ref{models}. It should be noted that in all cases less than
50$\%$ of the measured VLA flux density was recovered.  For the source
J0150+1308 (star-forming) only an extended model significantly larger
than the restoring beam produced a good fit. Fitting a point source
model on the map left residual emission on the dirty map, indicating a
more extended structure.  Thus the non-detection of J0150+1308
(star-forming) at 5~GHz is most likely due to over-resolution. On the
other hand, J0921+4509 (composite) could be fit equally well with a
point source, a relatively compact circular Gaussian ($\sim$
1.32~mas), or a more extended Gaussian source model (with size about
the same as the major axis of the restoring beam).

Brightness temperatures were determined for each source detected using
the formula
\begin{equation}
T_{\rmn{b,vlbi}}=1.22 \times 10^{12}(1+z){{S}\over{\theta_{\rmn{1}}\theta_{\rmn{2}}\nu^{2}}},
\end{equation}
where $S$ is the flux density (Jy), $\theta_1$ and $\theta_2$ are the
major and minor axes of the fitted Gaussian model component (mas or
FWHM for a point source) and $\nu$ is the observing frequency (GHz).

\begin{table}
  \caption{Modelfit results and brightness temperatures for all detections
    near or above 5$\sigma$. J0150+1308 (star-forming) could be fit well only by a
    circular-Gaussian model but J0921+4509 (composite) gave similarly good results
    when fit by a point source, a compact circular-Gaussian smaller
    than the beam, and a more extended model. Here we give the compact
    circular-Gaussian model size (see text). For the weakest detection,
    J1029+4829 (AGN), only a point model was used. In this case the geometric average of the beam major and minor
    axes is given as upper limit for the size.}
\label{models}
\begin{tabular}{ccccrr}
\hline
\hline
Source     & $\nu$ &   $S$   & Ratio   &$r$& $T_{\rm b}$\\
                  &  (GHz)  & ($\mu$Jy) & (EVN/VLA) & (mas) & ($10^6$K)\\
\hline
J0150+1308 & 1.658 & 456 & 0.304 & 59.3 & 0.10 \\
J0921+4509 & 1.658 & 111 & 0.079 & 1.32 & 34.9 \\
J1029+4829 & 4.990 & 126 & N/A & $<$9.4 & $>$0.12 \\
\hline
\hline
\end{tabular}
\end{table}

Because we could not distinguish between point and extended models in the
case of J0921+4509 (composite), we used a Monte Carlo simulation of
1000 trials to attempt to constrain the source size.  All Monte Carlo
simulations were carried out using ParselTongue
\citep[][]{kettenis06} to call the relevant AIPS task
repeatedly. Taking the 1.7~GHz field of J0213+1259 (composite) as a
non-detection field we used the AIPS task $uvsub$ to add a point
source directly to the $uv$-data. Random noise was simulated by
placing the source at a different location in the field with every
call to $uvsub$. The AIPS task $omfit$ was then used to fit a circular
Gaussian model to this point source. Trials were conducted with
initial modelfitting parameters of both 2~mas and 10~mas for the major
axis and giving the correct location of the simulated source.  In both
cases, the results were strongly peaked at a size of 0~mas though an
initial fit parameter of 10~mas occasionally led $omfit$ to give
results between 10~mas and 15~mas as the estimation of the size of our
point source. We repeated the trials adding phase errors of 10~rms by
using ParselTongue to manually edit the SN table of our dataset. When
fitting a 10~mas model to this data the secondary peak between 10~mas
and 15~mas was more pronounced.  So while one of our modelfit results
for J0921+4509 had a radius of up to 15.9~mas our simulations reveal
the radio source could still be unresolved with a much smaller radius,
especially given expected phase errors in our data.

Next we performed a similar trial but attempted to fit a Gaussian
model to an inserted Gaussian source of 10~mas. The initial model
provided to $omfit$ exactly matched the simulated source.  While
results of the modelfitting program tended to cluster around 10~mas,
there was a spread of $\pm$5~mas with and without phase errors of
10~rms. Thus any modelfit result obtained is likely accurate to within
$\pm$5~mas.  However, this trial never produced results in the region
of 1~mas.  In our simulations an extended model comparable to the
beamsize can be fit by a weak point source accurately, but it is not
possible to fit well an extended source with a compact model much
smaller than the beam. Since the VLBI structure of J0921+4509
(composite) can be fit well with compact models, it is most likely
that the true size of the source is about 1.32~mas, with an error of
$\pm 5$~mas.

To summarize, we conclude that we have made two safe detections, one
of which, J0150+1308 (star-forming), was resolved at 59.3~mas while
the other, J0921+4509 (composite), is practically unresolved at
1.32~mas. In the next section we will discuss the interpretation of
these results.

\begin{figure}
\includegraphics[width=\columnwidth]{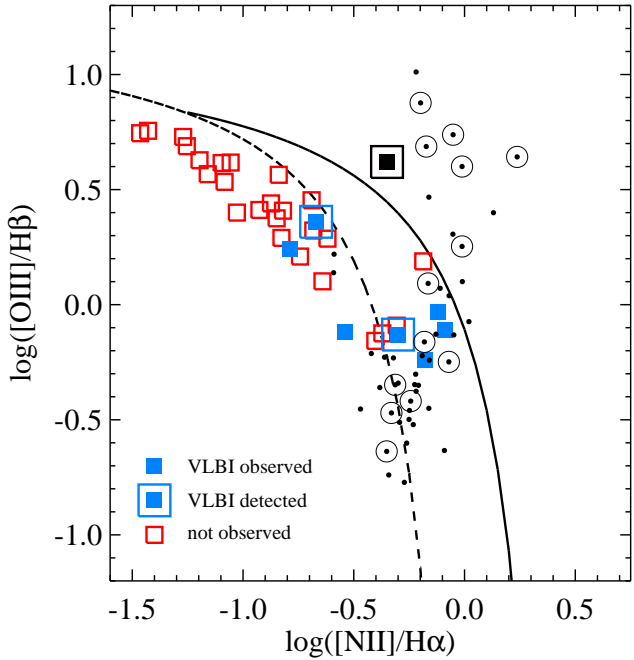}
\caption{\label{fig:bpt}Optical emission lines diagnostic diagram
  according to \citet[][BPT]{baldwin81}. The LBA sample studied in
  \citet{overzier09} is indicated by red/blue squares. Filled, blue squares
  mark LBAs that have been observed as part of our VLBI campaign,
  while (open) red squares mark LBAs that were not observed with
  VLBI. Objects detected during our EVN experiment are marked by the
  large, open blue squares. The black filled square indicates the
  AGN-like object J1029+4829 that was observed as part of this sample
  but is not an LBA.  Small dots indicate LIRGs from the Northern COLA
  sample from \citet{parra10} also observed with VLBI, with detections
  marked by the large open circles. Objects that lie in the region of
  the diagram in between the dotted and dashed lines are so-called
  starburst-AGN composite objects, while objects to the left of the
  dashed line and to the right of the solid line are HII-like and
  AGN-like objects, respectively \citep{kauffmann03,kewley06}. }
\end{figure}

\begin{figure}
\includegraphics[width=\columnwidth]{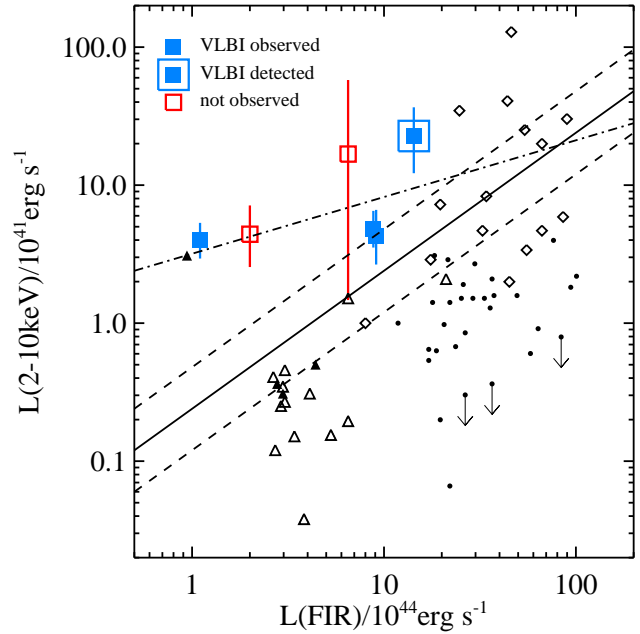}
\caption{\label{fig:lx} Hard X-ray luminosity vs. far infrared
  luminosity of LBAs from \citet{jia11}, who observed six composite
  LBAs with XMM-Newton. The four composite LBAs also
  observed with VLBI are shown in blue, while the two composite LBAs
  not observed with VLBI are marked red. The LBA composite J0921+4509
  that was detected during our VLBI experiment is marked with a large
  blue square. Other samples shown for comparison are as follows:
  LIRGs and LIRGs AGN from \citet{lehmer10} (open and filled
  triangles, resp.), LIRGs AGN (diamonds) and “Hard X-ray Quiet”
  galaxies (small points) from \citet{iwasawa09}. The solid line
  indicates the empirical $L_{X,2-10keV}-L_{FIR}$ relation from
  \citet{ranalli03}, with the dashed lines showing an associated
  uncertainty of a factor of two. The dash-dotted line is the best-fit
  relation for the composite LBAs found by \citet{jia11}.}
\end{figure}

\section{Discussion}

\subsection{Preexisting evidence for hidden AGN in composite LBAs}

In order to address the question of whether weak AGN are present in
the nuclear starburst regions of the LBAs, we specifically targeted
four LBAs having optical emission-line spectra intermediate between
pure starbursts and Type II AGN (J0213+1259, J0808+3948, J0921+4509,
J2103--0728). We also targeted three objects that, at least according
to their main optical emission line ratios, are not suspected of
having an obscured AGN present (J0150+1308, J0815+5004, J1353+6648),
and one source that is an AGN (J1029+4829).  These sources are shown
in the optical emission line diagnostics diagram in Figure
\ref{fig:bpt} (BPT diagram).

Previous work has shown that the LBAs that lie in the composite part
of the BPT diagram have a number of other properties as well that make
them peculiar with respect to the ``non-composite'' LBAs:
\citet{overzier09} showed that they are host to particularly dense and
compact starbursts (radii of order 100 pc), reaching stellar mass
densities comparable to those of dense star clusters, while being up
to 1000 times more massive (stellar masses of $\simeq10^{8}-10^9$
$M_\odot$).  They also tend to be the most massive galaxies in the LBA
sample, and have the highest SFR (several tens to a hundred $M_\odot$
yr$^{-1}$, the exception being J0808+3948). \citet{heckman11}
presented FUV spectra obtained with HST/COS that showed that these
dense, compact starburst regions drive powerful winds with median and
maximum velocities several times higher than that seen in typical
starbursts, including LBGs at high redshift. Based on the starburst
nature of the FUV spectra, the high wind speeds and the lack of
evidence of unobscured (i.e. Type I) AGN, it was argued that these are
momentum-driven winds caused by the intense supernova activity in the
compact starbursts at the base of the wind. Finally, \citet{jia11}
found the first plausible evidence for the presence of low luminosity,
obscured AGN in some or all of 6 composite LBAs observed in the
X-ray. The presence of an AGN was inferred from elevated levels
(factors of 2--10) of hard X-ray emission over that expected from the
SFR--$L_X$ sequence formed by pure star-forming galaxies (see Figure
\ref{fig:lx}). However, the SFR--$L_X$ correlation and its scatter are
not particularly well constrained, making the detection somewhat
ambiguous. A stacked spectrum showed the potential detection of the
6.4 keV iron emission line (which is a key feature of obscured AGN),
but only at the $\sim$2$\sigma$ level.  \citet{jia11} further found
that the ratios of the mid-infrared (24 $\mu$m) continuum to [O
III]$\lambda$5007 luminosities in these LBAs was higher than the
values for typical Type 2 AGN by an average of 0.8 dex. Combining all
these clues, it is likely that a hidden AGN is present in some or all
of these composite LBAs, but that their bolometric luminosities are
completely dominated by the intense starburst.

\subsection{New insights from the VLBI data}

The VLBI data on LBAs presented in this paper provide another piece of
information on the issue of AGN in LBAs.  As shown in Figure
\ref{fig:bpt}, we have detected one out of the four composite LBAs
observed with VLBI, and one out of the three HII-like LBAs. (Our third
detection of a known AGN will not be discussed further here).

On first sight, these results do not offer a clear dichotomy between
star-forming and composite LBAs. Because all LBAs have extended radio
structures due to star formation and supernova activity extending to
larger scales than those probed by VLBI \citep{basu07}, the emission
we have detected could in principle be coming from certain compact,
high surface brightness regions related to those structures. The fact
that three of the four composite LBAs do not show any VLBI radio
emission at all indicates that (1) the radio emission arising from
their star-forming regions is completely resolved at milliarcsecond or
parsec scales, and (2) no (bright) radio AGN is present in these
sources.

What other constraints do we have in order to study the relative
contributions from starburst and AGN activity in this sample?  Our
detections have brightness temperatures of $\sim10^5$ and $\sim10^7$ K
(see Table~\ref{models}).  These values are equal to (in the case of
J0150+1308) and well in excess of (J0921+4509) the typical upper-limit
for thermal emission arising from star-forming regions ($\sim10^5$ K),
thus indicating that the emission in these objects is likely of
non-thermal origin, at least in the case of J0921+4509
\citep{condon91}. Even though at these frequencies starburst emission
may be dominated by nonthermal emission as well, their maximum
brightness temperature still does not appear to exceed that limit. One
interpretation is that the observed radio emission is associated with
radio supernovae (RSNe) or supernova remnants (SNRs). The
non-detection at 5 GHz of both of our detections at 1.7~GHz supports
this conclusion: our sources have steep spectra which may be
indicative of RSNe or SNR.

\begin{table}
  \caption{Radio luminosity, physical size and resulting star formation rate for each of the 5 $\sigma$ detections.  Should the radio emission be from a starburst the radio luminosity can be used to estimate the star formation rate.  We calculated the SFR of our sources using the equation from \citet{basu07}.}
\label{physical}
\begin{tabular}{ccccc}
\hline
\hline
Source     & $L_{1.7}$ &   Physical Size   & SFR \\
&  ($10^{29}$ ergs s$^{-1}$ Hz$^{-1}$)  & (pc) & (M$_{\odot}$\,yr$^{-1}$) \\
\hline
J0150+1308 &  2.116 & 146.2 & 7.78\\
J0921+4509 & 1.388 & 4.76 & 5.10\\
\hline
J1029+4829 & 1.540 & $<$33.7 & 5.66\\
\hline
\hline
\end{tabular}
\end{table}

We can compare the total (non-thermal) radio luminosity detected from
J0150+1308, $L_{1.7~GHz}=2.1\times10^{29}$ erg s$^{-1}$ Hz$^{-1}$, and
J0921+4509, $L_{1.7~GHz}=1.4\times10^{29}$ erg s$^{-1}$ Hz$^{-1}$
(Table \ref{physical}), with that expected based on nearby
starburst galaxies for which it is possible to determine the relative
contributions of individual RSNe and SNRs. The brightest RSN observed
in starburst galaxies have a typical luminosity at 18 cm of the order
of $2\times10^{27}$ erg s$^{-1}$ Hz$^{-1}$, while the most luminous
RSN ever observed (SN1988z) had $2\times10^{28}$ erg s$^{-1}$
Hz$^{-1}$ \citep{bondi05}. It is thus clear that the radio emission
from LBAs requires large complexes of such SN (or an AGN). We will
employ the well-studied examples of such complexes observed in the
starburst galaxies Arp299-A \citep{perez09,perez10}, Arp220
\citep{lonsdale06,parra07}, and Mrk273 \citep{bondi05} to guide our
discussion.

The central $150\times85$ pc region of Arp299-A consists of a cluster
of 26 RSNe and SNRs detected at 5~GHz, with a total luminosity coming
from this region of $L_{5.0~GHz}=2\times10^{28}$ erg s$^{-1}$
Hz$^{-1}$, corresponding to $L_{1.7~GHz}\approx4.5\times10^{28}$ erg
s$^{-1}$ Hz$^{-1}$ assuming a median spectral index of -0.72 for the
region as a whole.  The Western nucleus of Arp220 is another example
of such a prolific supernova factory: it consists of 18 individual
RNSe/SNRs within a region of $50\times150$ pc and a total radio
luminosity of $L_{1.7~GHz}\approx1.6\times10^{29}$ erg s$^{-1}$
Hz$^{-1}$.\footnote{It should be noted that \citet{downes07} found evidence for a weak AGN in the West nucleus of Arp220.  If this is the case then the radio excess we detect would either be the result of many additional RNSe/SNRs, though that would be unlikely, or the result of a stronger AGN.} In order to minimize potential systematic biases when we
compare these objects with the LBAs, we will apply a scaling based on
the ratios of the physical areas ($A$) observed and the total IR
luminosities ($L_{IR}$) of these systems. The expected 18 cm
luminosities for the LBAs are then $f_{LBA} * L_R$, where
$f_{LBA}\approx(A_{LBA}/A_{Arp220/Arp299A})\times(L_{IR,LBA}/L_{IR,Arp220/Arp299A})$
with log$L_{IR,Arp220}=12.27$ and log$L_{IR,Arp299A}=11.53$. $f_{LBA}$
was typically in the range $1-2$, and was never larger than a factor
of 5, indicating that the comparison sample is well-matched to the
LBAs in terms of their SFRs and size of the physical regions
probed. Next, we calculate the ratio of observed to expected radio
luminosity for both LBAs.

For J0150+1308, the radio luminosity observed is, respectively,
$1.7\times$ and $2.8\times$ higher than that expected based on Arp220
and Arp299A.  For J0921+4509, and assuming that the source is resolved
(see Section 3), the observed luminosity is, respectively, $2.9\times$
and $4.8\times$ higher than that expected based on Arp220 and
Arp299A. If we assume that J0921+4509 is unresolved, its smaller area
leads to observed luminosities that are higher than expected by
factors of $4.7$ and $7.7$. Similarly, if we compare with the
brightest radio nucleus in Mrk273, N1 with
$L_{1.7~GHz}\approx7\times10^{28}$ erg s$^{-1}$ Hz$^{-1}$
\citep{bondi05}, the observed luminosities are again higher by factors
of 2--3 for both LBAs. To summarize, our comparison shows that the
radio luminosities of the LBAs are always higher than expected by
factor of 2--3 for J0150+1308 and a factor of 3--8 for J0921+4509
suggesting that more than just star formation may be contributing to
the radio emission observed.

\subsection{The starburst interpretation}

In \citet{overzier09} we argued that one explanation for the position
in the BPT diagram of composite LBAs such as J0921+4509, together with
their relatively low $L_{H\alpha}/L_{FIR}$, was that these are
post-starbursts where the O stars were deficient and ionization by
shocks (rather than photoionization by O stars) was very significant
(our so-called ``aged-starburst'' hypothesis, \S4.2.1 in that
paper). However, confirmation of shock heating in the form of enhanced
[SII] and [OI] line strengths was not found. New COS data on
J0921+4509 imply a young age of $\sim$3 Myr for the central starburst
(\citet{heckman11} and Borthakur et al., in prep.), ruling out that
model. These are young starbursts where the ionizing luminosity is at
least an order-of-magnitude larger than the rate of kinetic energy
supplied by supernovae. While this does not mean that shocks may not
be contributing (see \citet{heckman11} for evidence of high velocity
outflows), there is certainly no evidence for a weakness or an absence
of ionizing radiation from stars that would lead to shock-dominated
emission lines if this were an aged starburst. The starburst
interpretation also fails to explain the excess X-ray emission
observed in this source, unless an exotic starburst contribution to
the hard X-ray emission is invoked \citep{jia11}.

\subsection{The AGN interpretation}

A more likely interpretation is that (some) composite LBAs host an
AGN, consistent with the large excess compact radio emission observed
on VLBI scales (in the case of J0921+4509). Unfortunately, we do not
have sufficient resolution and sensitivity to look for further confirming evidence by
resolving the core-jet structure in the radio, and the non-detections
at 5~GHz imply that these low-luminosity
sources may have significantly steeper spectra than found
in radio-loud AGN \citep[e.g.][]{prandoni10, giroletti09}. On the other hand, there are now four
independent indications that an AGN may be present: (1) the composite
position in the BPT diagram, (2) the high VLBI luminosity, (3) the
compact radio size, and (4) the hard X-ray excess found by
\citet[][see also Fig. \ref{fig:lx}]{jia11}. The offset into the
AGN/composite region of the BPT diagram could then be the result of
the strong ionizing radiation from the starburst with a small
contribution from an obscured AGN (as well as some contribution from
shocks).

If we assume that the VLBI detection in J0921+4509 is due to a radio
AGN, we can compare it to typical radio-quiet quasars and Seyfert
galaxies. Taking $L_{X,2-10}=2.3\times10^{42}$ erg s$^{-1}$ from
\citet{jia11} and applying a factor of 2.86 to convert to
$L_{X,0.2-20}$ we find $L_R/L_X\approx1.3\times10^{-4}$, about
10$\times$ higher than inferred from the $L_R/L_X$ relation from
\citet[][]{laor08} who find $<L_R/L_X>\sim 10^{-5}$ (with a $1\sigma$
scatter of about a factor 3 in $L_R$) for radio-quiet quasars.  We
compare this source to the BH ``fundamental plane'' (FP) from
\citet[][M03]{merloni03}. Even though J0921+4509 was not detected at
5~GHz, we have extrapolated our 1.7~GHz detection to 5~GHz assuming a
spectral index of -0.1 (consistent with the (5$\sigma$) upper limit
for this source in the 5~GHz observation). At face value, the result
suggests a BH mass of a few times $10^7$ $M_\odot$. Much higher values
are unlikely given the constraint on the radio luminosity, although
there is significant uncertainty in the BH FP depending on the source
of the radio and X-ray emission and the accretion state
\citep[e.g. see][]{plotkin11}. In a recent paper, \citet{degasperin11}
find that a sample of low luminosity Type II AGN and LINERS do not lie
on the BH FP of \citet{merloni03}. Instead, based on their BH masses
inferred from stellar velocity dispersions, the M03 relation would
overpredict the $M_{BH}$ of their sources by up to two orders of
magnitude. In \citet{jia11} it was estimated that the candidate Type 2
AGN in the composite LBAs may have BH masses in the range of
$10^5-10^6$ $M_\odot$ if these objects are radiating at the Eddington
limit, but they could be higher for sub-Eddington BHs.  On the other hand, \citet{overzier09} showed that the central mass concentration on a scale of $\sim$100 pc is consistent with a ``proto-bulge'' of order $10^9$ $M_\odot$ in stars. Assuming a standard bulge-black hole mass scaling relation, it is hard to imagine the presence of a black hole substantially more massive than $\sim10^7$ $M_\odot$.  The presence of
a compact radio core has at least made it convincingly clear that an
AGN is present in J0921+4509, but it is impossible to give better
constraints on the BH mass than $10^5$--$10^7$ $M_\odot$.

We can compare the results from our small sample with the results of
\citet{parra10} who analyzed the radio AGN fraction among a much
larger sample of relatively nearby, ``warm'' IR-luminous galaxies
(wLIRGs) across the BPT diagram, also using VLBI. In some respects,
their sample can be considered to be the more obscured equivalent of
our UV-selected LBA sample \citep[see][]{overzier09}.  The wLIRGs and
LBAs are shown in Figure \ref{fig:bpt} (small dots), where sources
with large circles indicate VLBI detections. While most of the wLIRG
sources with optical spectra indicative of a dominant AGN were
detected in VLBI, most ($\sim80$\%) of the sources in the
AGN-starburst composite region of the diagram were, in fact, not \citep[see also][]{corbett03}. This
situation is similar to the non-detections for three of our four
composite LBAs. One may naively expect that the chance of detecting an
AGN increases significantly as one moves from pure HII-like objects to
pure AGN-like objects across the BPT diagram, contrary to what is
observed.  One explanation for the lack of VLBI detections among the
composite LBAs and wLIRGs could be that the high gas columns of the
dense nuclear gas associated with the starbursts leads to significant
free-free absorption, which would especially affect these composite
systems. In this case, it would be better to observe at higher
frequencies in order to offset the $\lambda^2$ dependence of the
free-free opacity, although we note that we did not detect anything at
5.0~GHz either (including the known 1.7~GHz source J0921+4509).  On
the other hand, even with a relatively high $L_R/L_X$ of $10^{-4}$ as
found for J0921+4509, the remaining composite LBAs sources would have
been hard to detect given their hard X-ray luminosities (about a
factor of 10 lower). Not only are the radio AGN expected to be
intrinsically faint, the relatively high redshift of our sources
($z\sim0.2$) complicates their detectability with VLBI.
Alternatively, given AGN variability as observed in some LLAGN
\citep{bell11}, or the episodic accretion expected in some
cosmological galaxy formation models \citep[e.g.][]{hopkins10} it is
perhaps not surprising that our very small sample observed produced
only one clear detection of an accreting BH in the radio as the X-ray
observations were performed at a different epoch.

\section{Summary and future prospects}

This paper presents the results of a pilot program to observe LBAs
with composite spectra intermediate between pure starbursts and Type
II AGN for comparison with objects whose optical emission-line ratios did not
appear to suggest the presence of an obscured AGN.  The
VLBI data obtained with the EVN were combined with data previously
obtained in the X-ray from XMM-Newton, optical from SDSS and mid- and
far- IR from Spitzer to further constrain the nature of these objects
that are of special interest given their demonstrated similarities to
galaxies that were common only at high redshift.  While it might be expected
that composite objects are detected more frequently than pure HII-like regions only $\sim20$\%
of our composite LBAs were detected, possibly due to free-free absorption.  

Three sources, J0150+1308, J0921+4509, and J1029+4829 were detected
above 5$\sigma$, and all had brightness temperatures close to or in
excess of the typical upper-limit for thermal emission from
star-forming regions.  In addition, J0150+1308 and J0921+4509 had
radio luminosities 2-8 times above that expected from multiple radio
supernovae and supernova remnants in the starbursts Arp220, Arp229A
and Mrk273 after a careful normalization based on the different
resolutions and total SFRs.

In particular, the composite LBA J0921+4509 displayed a compact core, most
likely unresolved as confirmed by Monte Carlo simulation,
emitting around 10$\%$ of observed VLA flux.  The source also
possessed an excess $L_R$ expected from the $L_R$/$L_X$ ratio of
radio-quiet AGN, while its $L_X$ was found to be high relative
to the value expected based on its SFR \citep{jia11}. This suggests a
source of radio and X-ray emission beyond that expected of a pure
starburst. We conclude therefore that this excess is most probably due
to an obscured AGN with an estimated BH mass of $\sim10^{5-7}$ $M_{\odot}$.

As stated in \S1, one of our main motivations for probing BH formation
in these LBAs as local analogs of LBGs is the question of whether the young starburst-dominated
galaxies seen at high redshift already carry the ``seeds'' of
supermassive BHs detected in the local universe. If so, these
small-to-moderate mass BHs at high redshift may reveal their presence
through AGN activity.  It is currently not clear
whether BHs would be able to accrete very efficiently in these young
and dense starburst dominated regions, or whether a delay should exist
between the onset of a central starburst and the accretion onto the
central BH \citep[e.g.][]{norman88,davies07}. Taken together, our
X-ray and radio results shown here indicate that LLAGN can co-exist in
young, dense starbursts, but our sample is currently too small to put
strong constraints on this timing argument.

It is important to point out that at radio wavelengths, the
contribution of these AGN is still only a fraction of the total radio
flux density, making the identification of such sources based on radio
data practically impossible much beyond $z\sim0.2$ (see below). Even
in the X-rays, the combined luminosity from SFR and candidate AGN in
LBAs \citep[][and Fig. 3]{jia11} is lower or comparable to that
detected in the deepest X-ray stacks of star-forming galaxies at
$z\sim3-4$ \citep{laird06,zheng10}, illustrating that individual
detections are likely out of reach for current X-ray
observatories. Also discussed in this paper, the interpretation of
optical emission line diagnostic diagrams for identifying weak AGN is
not trivial either, especially in the case of shocks or due to the
limited physical resolution achieved in high redshift data \citep[see
Fig. 12 in][for a large compilation of BPT information from low and
high redshift galaxies]{overzier09}. These limitations indicate the
important role that nearby galaxies still play in studying the
AGN-starburst (or BH-host galaxy) connection.

\begin{figure}
\includegraphics[width=\columnwidth]{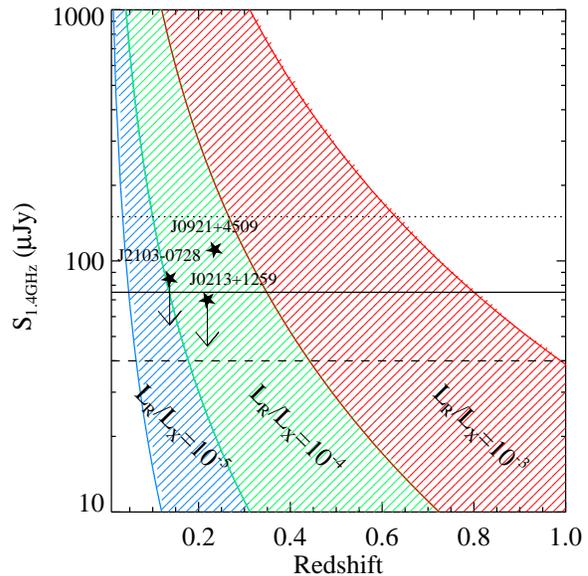}
\caption{\label{fig:sens}Predicted flux densities for low luminosity
  radio AGN as a function of redshift. The blue, green and red shaded
  areas show the 1.4 GHz flux densities expected for a compact core
  having $L_R/L_X$ ratios of $10^{-5}$ (i.e. typical radio-quiet AGN),
  $10^{-4}$ (i.e. 10$\times$ more luminous), and $10^{-3}$
  (i.e. 100$\times$ more luminous). The upper boundary of each region
  assumes $L_X\sim5\times10^{42}$ erg s$^{-1}$ (i.e. comparable to
  that of J0921+4509), while the lower boundaries are for sources that
  are a factor 10 less luminous in the X-rays (i.e. comparable to the
  other composite LBAs studied in this paper). Our detection of the
  probable AGN in J0921+4509 is indicated by the star, while two
  undetected composite LBAs are shown as upper limits. Horizontal
  lines indicate typical EVN 5$\sigma$ sensitivities achieved in a 1
  hr (dotted line) and 4 hr observation (solid line) both at 1 Gbps
  recording, as well as the projected 4h sensitivity of the EVN with new
   telescopes becoming available in the next few years (dashed line).}
\end{figure}

Our pilot program of searching for AGN in the most extreme UV-selected
starbursts in the relatively nearby universe demonstrates the unique
power of VLBI to assess the prevalence of accreting BHs and SN
complexes in young, star-forming galaxies. With this goal in mind, we
conclude with investigating the prospects for future VLBI detection
experiments.  In Fig. \ref{fig:sens} we show the predicted 1.4 GHz
flux densities for (weak) radio AGN as a function of redshift up to
$z=1$.  The blue, green and red shaded areas show the 1.4 GHz flux
densities expected for a compact core having $L_R/L_X$ ratios of
$10^{-5}$ (i.e. typical radio-quiet AGN), $10^{-4}$ (i.e. that
measured for the LBA composite J0921+4509), and $10^{-3}$. The upper
boundary of each region assumes $L_X\sim5\times10^{42}$ erg s$^{-1}$
(i.e. comparable to that of J0921+4509), while the lower boundaries
are for sources that are a factor 10 less luminous in the X-ray
(i.e. comparable to the other composite LBAs studied in this
paper). Horizontal lines indicate typical EVN 5$\sigma$ sensitivities
achieved in a 1~hr (dotted line) and 4~hr observation (solid line)
both at 1 Gbps recording, as well as the projected 4~hr sensitivity 
expected in the next few years (dashed line). A
source similar to J0921+4509 both in $L_X$ and $L_R/L_X$ can be
detected to a redshift of $z\sim0.4$, while it could be seen out to
$z\sim0.8$ if its radio-loudness ($L_R/L_X$) was $10\times$ higher. In
contrast, the prospects for sources that are radio-quiet and a factor
ten less luminous in the X-ray are grim: such sources could
potentially be detected only out to $z\sim0.1$.

The recently increased sky coverage in the UV (GALEX), optical (SDSS) and IR (WISE) will allow the efficient selection of exotic starbursts, composites, and AGN at relatively low redshifts suitable for VLBI. A systematic survey of their pc-scale radio emission offers a unique chance of gaining new insights into the complex nuclear processes occurring in galaxies over a very wide redshift range.

\section*{Acknowledgments}

The EVN is a joint facility of European, Chinese, South African and
other radio astronomy institutes funded by their national research
councils. We thank Patricia Ar\'evalo, Jenny Greene, Lisa Kewley, and
Miguel P\'erez Torres for discussions.  We would also like to thank the referee for his or her helpful feedback.

\label{lastpage}


\begin{thebibliography}{99}
\bibitem[Adelberger et al.(2005)]{adelberger05} Adelberger, K.~L., Steidel, C.~C., Pettini, M., Shapley, A.~E., Reddy, N.~A., \& Erb, D.~K.\ 2005, \apj, 619, 697 
\bibitem[Alexander et al.(2005)]{alexander05} Alexander, D.~M., Bauer, F.~E., Chapman, S.~C., Smail, I., Blain, A.~W., Brandt, W.~N., \& Ivison, R.~J.\ 2005, \apj, 632, 736 
\bibitem[Baldwin et al.(1981)]{baldwin81} Baldwin, J.~A., Phillips, M.~M., \& Terlevich, R.\ 1981, \pasp, 93, 5 
\bibitem[Basu-Zych et al.(2007)]{basu07} Basu-Zych, A.~R., et al.\ 2007, \apjs, 173, 457 
\bibitem[Basu-Zych et al.(2009)]{basu09} Basu-Zych, A., et al.\ 2009, \apjl, 699, L118
\bibitem[Bell et al.(2011)]{bell11} Bell, M.~E., Tzioumis, T., Uttley, P., et al.\ 2011, \mnras, 411, 402 
\bibitem[Bondi et al.(2005)]{bondi05} Bondi, M., P{\'e}rez-Torres, M.-A., Dallacasa, D., \& Muxlow, T.~W.~B.\ 2005, \mnras, 361, 748 
\bibitem[Biggs et al.(2010)]{biggs10} Biggs, A. D., et al.\ 2010, \mnras, 408, 342
\bibitem[Condon et al.(1991)]{condon91} Condon, J.~J., Huang, Z.~P., Yin, G.~F., \& Thuan, T.~X. \ 1991,\apj, 378, 65C
\bibitem[Corbett et al.(2003)]{corbett03} Corbett, E.~A., Kewley, L. Appleton, P.~N., Charmandaris, V., Dopita, M.~A., Heisler, C.~A., Norris, R.~P., Zezas, A., \& Marston, A. \ 2003, \apj, 583, 670 
\bibitem[Davies et al.(2007)]{davies07} Davies, R.~I., S{\'a}nchez, F.~M., Genzel, R., Tacconi, L.~J., Hicks, E.~K.~S., Friedrich, S., \& Sternberg, A.\ 2007, \apj, 671, 1388 
\bibitem[Downes \& Eckart(2007)]{downes07} Downes, D., \& Eckart, A. \ 2007, \aap, 468L, 57D
\bibitem[de Gasperin et al.(2011)]{degasperin11} de Gasperin, F., Merloni, A., Sell, P., Best, P., Heinz, S., \& Kauffmann, G.\ 2011, \mnras, 835 
\bibitem[Diamond (1995)]{diamond95} Diamond, P.J. 1995, in ASP Conf. Ser. 82, Very Long Baseline Interferometry and the VLBA, ed. J.A. Zensus, P.J. Diamond, \& P.J. Napier, 227
\bibitem[Di Matteo et al.(2005)]{dimatteo05} Di Matteo, T., Springel, V., \& Hernquist, L.\ 2005, \nat, 433, 604 
\bibitem[Elmegreen et al.(2008)]{elmegreen08} Elmegreen, B.~G., Bournaud, F., \& Elmegreen, D.~M.\ 2008, \apj, 684, 829 
\bibitem[Gallimore et al.(2006)]{gallimore06} Gallimore, J.~F., Axon, D.~J., O'Dea, C.~P., Baum, S.~A., \& Pedlar, A.\ 2006, \aj, 132, 546 
\bibitem[Gebhardt et al.(2000)]{gebhardt00} Gebhardt, K., et al.\ 2000, \apjl, 539, L13 
\bibitem[Giroletti \& Panessa(2009)]{giroletti09} Giroletti, M., \& Panessa, F.\ 2009 \apjl, 706, 260
\bibitem[Gon\c calves et al.(2010)]{goncalves10} Gon\c calves, T.~S., et al.\ 2010, \apj, 724,1373G
\bibitem[Heckman et al.(2005a)]{heckman05a} Heckman, T.~M., Ptak, A., Hornschemeier, A., \& Kauffmann, G.\ 2005a, \apj, 634, 161 
\bibitem[Heckman et al.(2005b)]{heckman05b} Heckman, T.~M., et al.\ 2005b, \apjl, 619, L35 
\bibitem[Heckman et al.(2011)]{heckman11} Heckman, T.~M., et al.\ 2011, \apj, 730, 5 
\bibitem[Hoopes et al.(2007)]{hoopes07} Hoopes, C.~G., Heckman, T.~M., Salim, S., et al.\ 2007, \apjs, 173, 441 
\bibitem[Hopkins \& Quataert(2010)]{hopkins10} Hopkins, P.~F., \& Quataert, E.\ 2010, \mnras, 407, 1529 
\bibitem[Iwasawa et al.(2009)]{iwasawa09} Iwasawa, K., Sanders, D.~B., Evans, A.~S., Mazzarella, J.~M., Armus, L., \& Surace, J.~A.\ 2009, \apjl, 695, L103 
\bibitem[Jia et al.(2011)]{jia11} Jia, J., Ptak, A., Heckman, T.~M., Overzier, R.~A., Hornschemeier, A., \& LaMassa, S.~M.\ 2011, \apj, 731, 55 
\bibitem[Kauffmann et al.(2003)]{kauffmann03} Kauffmann, G., et al.\ 2003, \mnras, 346, 1055 
\bibitem[Kettenis et al. (2006)]{kettenis06} Kettenis, M., van Langevelde, H.~J., Reynolds, C., Cotton, B.\ 2006, ASP Conference Series 351, 497
\bibitem[Kewley et al.(2006)]{kewley06} Kewley, L.~J., Groves, B., Kauffmann, G., \& Heckman, T.\ 2006, \mnras, 372, 961 
\bibitem[Kewley et al.(2000)]{kewley00} Kewley, L.~J., Heisler, C.~A., Dopita, M.~A., \& Sutherland, R.\ 2000, \apj, 530, 704
\bibitem[Kimball et al.(2011)]{kimball11} Kimball, Amy E., Kellermann, K.~I., Condon, J.~J., Ivezi\'{c}, \u{Zeljiko}, \& Perley, Richard A.\ 2011, \apj, 739L, 29K
\bibitem[Laor \& Behar (2008)]{laor08} Laor, A., \& Behar, E. \ 2008, \mnras, 390, 847
\bibitem[Laird et al.(2006)]{laird06} Laird, E.~S., Nandra, K., Hobbs, A., \& Steidel, C.~C.\ 2006, \mnras, 373, 217 
\bibitem[Lehmer et al.(2005)]{lehmer05} Lehmer, B.~D.,  Brandt, W.~N, Alexander, D.~M, Bauer, F.~E, Conselice, C.~J, Dickinson, M.~E, Giavalisco, M, Grogin, N.~A, Koekemoer, A.~M, Lee, K.~S, Moustakas, L.~A, \& Schneider, D.~P\ 2005, \aj, 129, 1L
\bibitem[Lehmer et al.(2010)]{lehmer10} Lehmer, B.~D., Alexander, D.~M., Bauer, F.~E., Brandt, W.~N., Goulding, A.~D., Jenkins, L.~P., Ptak, A., \& Roberts, T.~P.\ 2010, \apj, 724, 559 
\bibitem[Lonsdale et al.(2006)]{lonsdale06} Lonsdale, C.~J., Diamond, P.~J., Thrall, H., Smith, H.~E., \& Lonsdale, C.~J.\ 2006, \apj, 647, 185 
\bibitem[Magorrian et al.(1998)]{magorrian98} Magorrian, J., et al.\ 1998, \aj, 115, 2285 
\bibitem[Merloni et al.(2003)]{merloni03} Merloni, A., Heinz, S., \& di Matteo, T.\ 2003, \mnras, 345, 1057 
\bibitem[Norman \& Scoville(1988)]{norman88} Norman, C., \& Scoville, N.\ 1988, \apj, 332, 124 
\bibitem[Ouchi et al.(2008)]{ouchi08} Ouchi, M., et al.\ 2008, \apjs, 176, 301 
\bibitem[Overzier et al.(2008)]{overzier08} Overzier, R.~A., et al.\ 2008, \apj, 677, 37 
\bibitem[Overzier et al.(2009)]{overzier09} Overzier, R.~A., et al.\ 2009, \apj, 706, 203 
\bibitem[Overzier et al.(2010)]{overzier10} Overzier, R.~A., Heckman, T.~M., Schiminovich, D., Basu-Zych, A., Gon{\c c}alves, T., Martin, D.~C., \& Rich, R.~M.\ 2010, \apj, 710, 979 
\bibitem[Overzier et al.(2011)]{overzier11} Overzier, R.~A., et al.\ 2011, \apjl, 726, L7 
\bibitem[Padovani et al.(2011)]{padovani11} Padovani, P., Miller, N., Kellermann, K.~I., Mainieri, V., Rosati, P., \& Tozzi, P.\ 2011, \apj, 740, 20P
\bibitem[Parra et al.(2007)]{parra07} Parra, R., Conway, J.~E., Diamond, P.~J., Thrall, H., Lonsdale, C.~J., Lonsdale, C.~J., 
\& Smith, H.~E.\ 2007, \apj, 659, 314 
\bibitem[Parra et al.(2010)]{parra10} Parra, R., Conway, J.~E., Aalto, S., Appleton, P.~N., Norris, R.~P., Pihlstr{\"o}m, Y.~M., \& Kewley, L.~J.\ 2010, \apj, 720, 555 
\bibitem[Perez-Torres et al.(2009)]{perez09} Perez-Torres, Romero-Canizales, C., M.~A., Alberdi, A.,  \& Polatidis, A.\ 2009,\aap, 507, L17-L20 
\bibitem[Perez-Torres et al.(2010)]{perez10} Perez-Torres, M.~A., Alberdi, A., Romero-Canizales, C., \& Bondi, M.\ 2010, \aap, 519, L5 
\bibitem[Plotkin et al.(2011)]{plotkin11} Plotkin, R.~M., Markoff, S., Kelly, B.~C., Koerding, E., \& Anderson, S.~F.\ 2011, arXiv:1105.3211 
\bibitem[Prandoni et al.(2010)]{prandoni10} Prandoni, I., de Ruiter, H.~R., Parma, P. Gregorini, L., \& Ekers, R.~D.\ 2010, \aap, 510, 42 
\bibitem[Ranalli et al.(2003)]{ranalli03} Ranalli, P., Comastri, A., \& Setti, G.\ 2003, \aap, 399, 39 
\bibitem[Rich et al.(2011)]{rich11} Rich, J.~A., Kewley, L.~J., \& Dopita, M.~A.\ 2011, \apj, 734, 87 
\bibitem[Shepherd et al. (1994)]{shepherd94} Shepherd, M.~C., Pearson, T.~J. \& Taylor, G.~B.\ 1994,BAAS, 26, 987
\bibitem[Steidel et al.(2002)]{steidel02} Steidel, C.~C., Hunt, M.~P., Shapley, A.~E., Adelberger, K.~L., Pettini, M., Dickinson, M., \& Giavalisco, M.\ 2002, \apj, 576, 653 
\bibitem[Terashima \& Wilson(2003)]{terashima03} Terashima, Y., \& Wilson, A.~S.\ 2003, \apj, 583, 145 
\bibitem[Yuan et al.(2010)]{yuan10} Yuan, T.-T., Kewley, L.~J., \& Sanders, D.~B.\ 2010, \apj, 709, 884 
\bibitem[Zheng et al.(2010)]{zheng10} Zheng, Z.~Y., Wang, J.~X., Finkelstein, S.~L., Malhotra, S., Rhoads, J.~E., \& Finkelstein, K.~D.\ 2010, \apj, 718, 52 
\end{thebibliography}
\end{document}